\documentclass[english,a4paper,aip,pof,reprint,amsmath,amssymb,superscriptaddress,nofootinbib]{revtex4-1}
\usepackage{preamble}
\graphicspath{{Figures/},{Figures/SEM/}}

\usepackage{pdfpages}
\makeatletter
\AtBeginDocument{\let\LS@rot\@undefined}
\makeatother
\AtEndDocument{
  \foreach \x in {1,...,10}
  {%
    \includepdf[pages={\x,{}}]{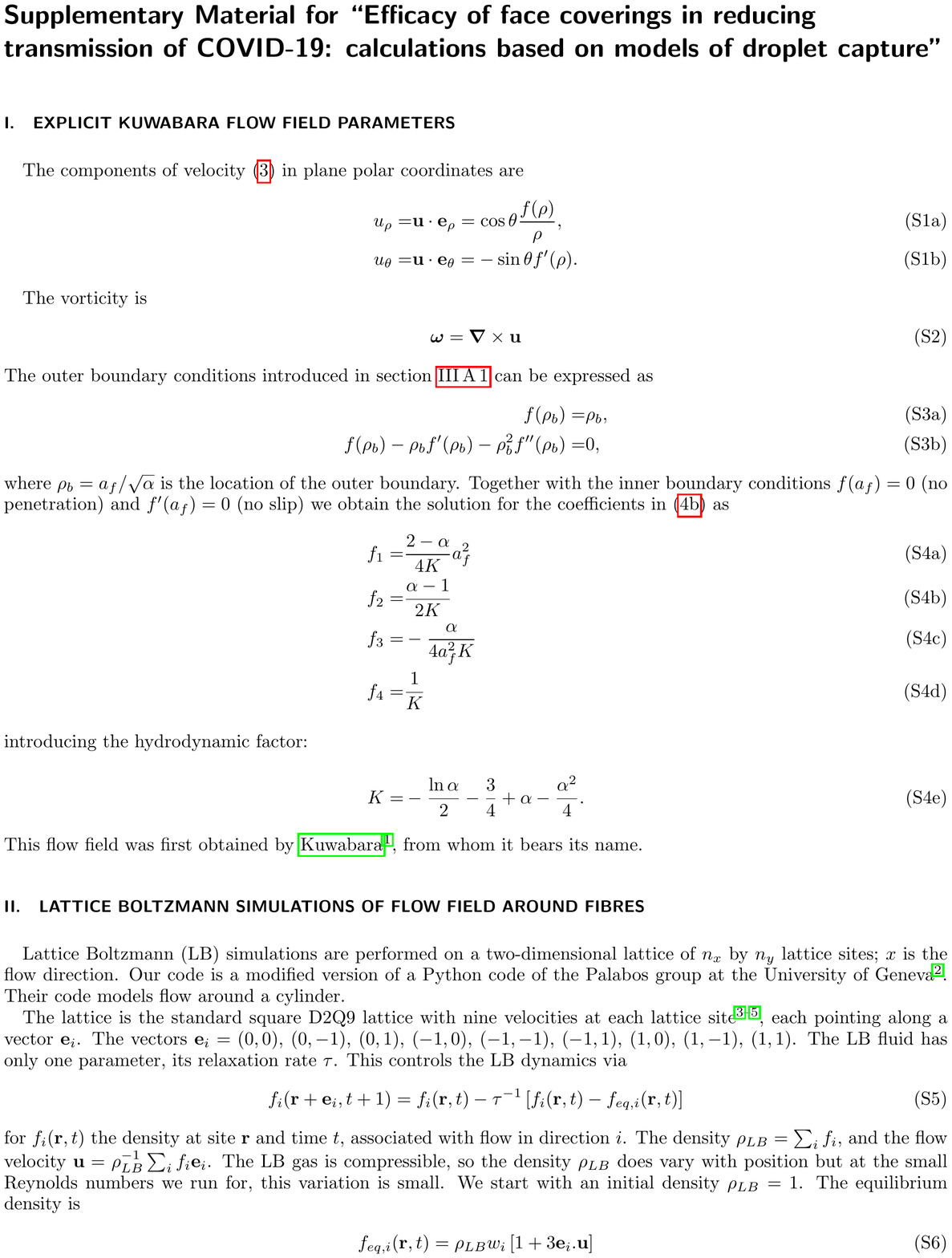}
  }
}

\usepackage{xr}
\begin{document}

\title{Efficacy of face coverings in reducing transmission of COVID-19: calculations based on models of droplet capture}
\date{\today}

\author{Joshua F. Robinson}
\email{joshua.robinson@bristol.ac.uk}
\affiliation{H.\ H.\ Wills Physics Laboratory, University of Bristol, Bristol, BS8 1TL, United Kingdom}
\author{Ioatzin Rios de Anda}
\affiliation{H.\ H.\ Wills Physics Laboratory, University of Bristol, Bristol, BS8 1TL, United Kingdom}
\affiliation{School of Mathematics, University Walk, University of Bristol, BS8 1TW, United Kingdom}
\author{Fergus J. Moore}
\affiliation{H.\ H.\ Wills Physics Laboratory, University of Bristol, Bristol, BS8 1TL, United Kingdom}
\affiliation{School of Mathematics, University Walk, University of Bristol, BS8 1TW, United Kingdom}
\author{Jonathan P. Reid}
\affiliation{School of Chemistry, Cantock's Close, University of Bristol, Bristol, BS8 1TS, United Kingdom}
\author{Richard P. Sear}
\affiliation{Department of Physics, University of Surrey, Guildford, GU2 7XH, United Kingdom}
\author{C. Patrick Royall}
\affiliation{Gulliver UMR CNRS 7083, ESPCI Paris, Universit\'e PSL, 75005 Paris, France}
\affiliation{H.\ H.\ Wills Physics Laboratory, University of Bristol, Bristol, BS8 1TL, United Kingdom}
\affiliation{School of Chemistry, Cantock's Close, University of Bristol, Bristol, BS8 1TS, United Kingdom}
\affiliation{Centre for Nanoscience and Quantum Information, University of Bristol, Bristol, BS8 1FD, United Kingdom}

\begin{abstract}
  In the COVID--19 pandemic, among the more controversial issues is the use of masks and face coverings.
  Much of the concern boils down to the question -- \emph{just how effective are face coverings?}
  One means to address this question is to review our understanding of the physical mechanisms by which masks and coverings operate -- \emph{steric interception, inertial impaction, diffusion} and \emph{electrostatic capture}.
  We enquire as to what extent these can be used to predict the efficacy of coverings.
  We combine the predictions of the models of these mechanisms which exist in the filtration literature and compare the predictions with recent experiments and lattice Boltzmann simulations, and find reasonable agreement with the former and good agreement with the latter.
  Building on these results, we explore the parameter space for woven cotton fabrics to show that three-layered cloth masks can be constructed with comparable filtration performance to surgical masks under ideal conditions.
  Reusable cloth masks thus present an environmentally friendly alternative to surgical masks so long as the face seal is adequate enough to minimise leakage.
\end{abstract}

\maketitle

\section{Introduction}

Face coverings have become a common (though controversial) motif
of the global response to the COVID--19 pandemic \cite{howard2020,delve2020,greenhalgh2020,prather2020}.
At the time of writing, 139 countries have mandated the use of face coverings (or already practiced universal masking) in public spaces such as on public transport, 19 countries mandate coverings on a regional level and a further 17 countries recommend (but do not require) their use \cite{masks4allCountries}.
The World Health Organisation has recently reversed their earlier policy on face coverings, and now advise that the public wear them and offer some guidance on the essential features of effective coverings \cite{whoMaskAdvice}.

SARS-CoV-2 is transmitted primarily by the \emph{airborne route}, \ie by direct inhalation of aerosolised particles containing virus \cite{morawska2020,anderson2020,vandoremalen2020,fears2020,lednicky2020,jayaweera2020,kampf2020,anderson2020,azimuddin2020,asadi2020}.
Face coverings work to prevent this transmission route by suppressing onwards transmission of the virus on exhalation \cite{cheng2020} (so-called ``source control'') or to provide protection to the wearer on inhalation \ie as personal protective equipment (PPE).
The former is especially important in this pandemic because the majority of cases of transmission seem to occur from asymptomatic or presymptomatic patients \cite{moghadas2020,vuorinen2020,sakurai2020,arons2020,treibel2020,streeck2020,ferretti2020,james2020,emery2020,prather2020,adam2020}.
Following the emergence of more infectious variants of SARS-CoV-2, some policy makers have mandated the wearing of medical-grade PPE in public spaces \cite{bavariaFFP2Mandate}.

The literature on face coverings is limited \cite{cowling2010,delve2020}, and there is a great deal of inconsistency and a lack of clarity in the guidance concerning their use.
The academic literature is a combination of medical studies (using either live wearers \cite{cowling2009,bischoff2011,aiello2012} or mannequins \cite{balazy2006,dato2006,ai2018}), retrospective studies \cite{delve2020,wong2004,wang2020,chu2020,zeng2020}, epidemiological modelling \cite{delve2020,tian2020,jombart2020,stutt2020} engineering studies (particularly in the filtration literature) \cite{dato2006,rengasamy2010,wang2013,davies2013,ai2018,zangmeister2020,konda2020,lustig2020,wangData2020} and aerosol science \cite{hinds1999,setti2020,vuorinen2020,vandoremalen2020,han2020,wang2020,morawska2020,prather2020}.
Such a complex phenomenon as airborne transmission depends on very many parameters (\eg air flow, humidity, separation, mask fit).
The disparate disciplines which have considered the use of face coverings take wildly differing approaches, and there seems to be a lack of any consistent experimental protocol, and studies typically only address a subset of the parameters upon which transmission depends.

The mechanisms by which droplets\footnote{
Here we use \emph{droplet} to refer to liquid particles of any size, independent of the mechanisms by which they transmit pathogens.
In the literature the terms \emph{droplet} and \emph{aerosol} are used to describe distinct routes for disease transmission, mediated by liquid droplets in different size regimes.
There is a great deal of ambiguity involved in distinguishing these two regimes, which is discussed in more detail in \Refcite{vuorinen2020}.
To avoid confusion and an arbitrary classification into different size regimes, we use a single term for particles of all sizes.}
are captured by filters are reasonably well-established \cite{wang2001}.
There are four principle mechanisms by which droplets may be captured by fibres in a covering which concern us here \cite{wang2013}.
\begin{itemize}
  \item \emph{Steric interception} -- capture neglecting inertia, so a droplet follows stream lines of the air but collides with a fibre due to the size of the droplet.
  \item \emph{Inertial impaction} -- where inertia is taken into account resulting in the droplet deviating from stream lines and colliding with the fibre.
  \item \emph{Diffusion} -- diffusion of droplets in the air leads to contact with a fibre.
  \item \emph{Electrostatic capture} -- Coulombic and/or dipolar attractions between the droplets and fibres pull the droplet into contact.
  Note that the previous three mechanisms assume no interaction until particle/fibre contact.
  Studying this mechanism requires knowledge of the charge distribution in the droplets and fibres.
\end{itemize}
Gravitation can also play a role in droplet capture, however this is negligible compared to the other mechanisms outlined above \cite{chen1993}.
The filtration literature's focus on these mechanisms was primarily motivated by developing medical grade PPE.
However, experimental work during the pandemic has confirmed the potential of household fabrics to effectively filter some virus-bearing particles \cite{zangmeister2020,konda2020,lustig2020,wangData2020}.

Here we shall primarily focus on those filtration mechanisms pertinent to droplet capture in cloth masks: interception and inertial impaction.
We review the literature which addresses these mechanisms and assess experimental measurements of droplet capture by face coverings.
We give a technical account of filtration theory in a rigorous fashion by borrowing some ideas from soft matter physics.
By clearly articulating its underlying assumptions, we are able to extend the standard theory to begin to treat household fabrics.
Our work provides a route through which mask design can be optimised, and further questions of public policy can be explored in future \eg the importance of mask fit.

Our model predicts that multi-layered masks made from standard household fabrics should provide comparable filtration performance to surgical masks under ideal conditions, though practical mask performance crucially depends on the fit.
We conclude that for many three-layered cloth masks capture of droplets larger than $\gtrsim \SI{3}{\um}$ is highly effective.
For smaller (\SIrange{0.1}{3}{\um}) droplets, the efficacy is dependent on the type of material from which the face covering is comprised, but some materials can achieve excellent protection ($\ge \SI{95}{\percent}$) for $\gtrsim \SI{1}{\um}$ droplets which is comparable to surgical masks.

This paper is organised as follows: in section \ref{sec:materials} we describe experiments exploring the material properties of fabrics.
Section \ref{sec:single-fibres} is dedicated to theory and simulations for filtration by a single-fibre.
Then in section \ref{sec:total-filter} we investigate the filtration properties of fabrics by combining the work of previous sections.
We discuss the significance of our findings and conclude in section \ref{sec:conclusion}.

\begin{figure*}
  \includegraphics[width=0.9\linewidth]{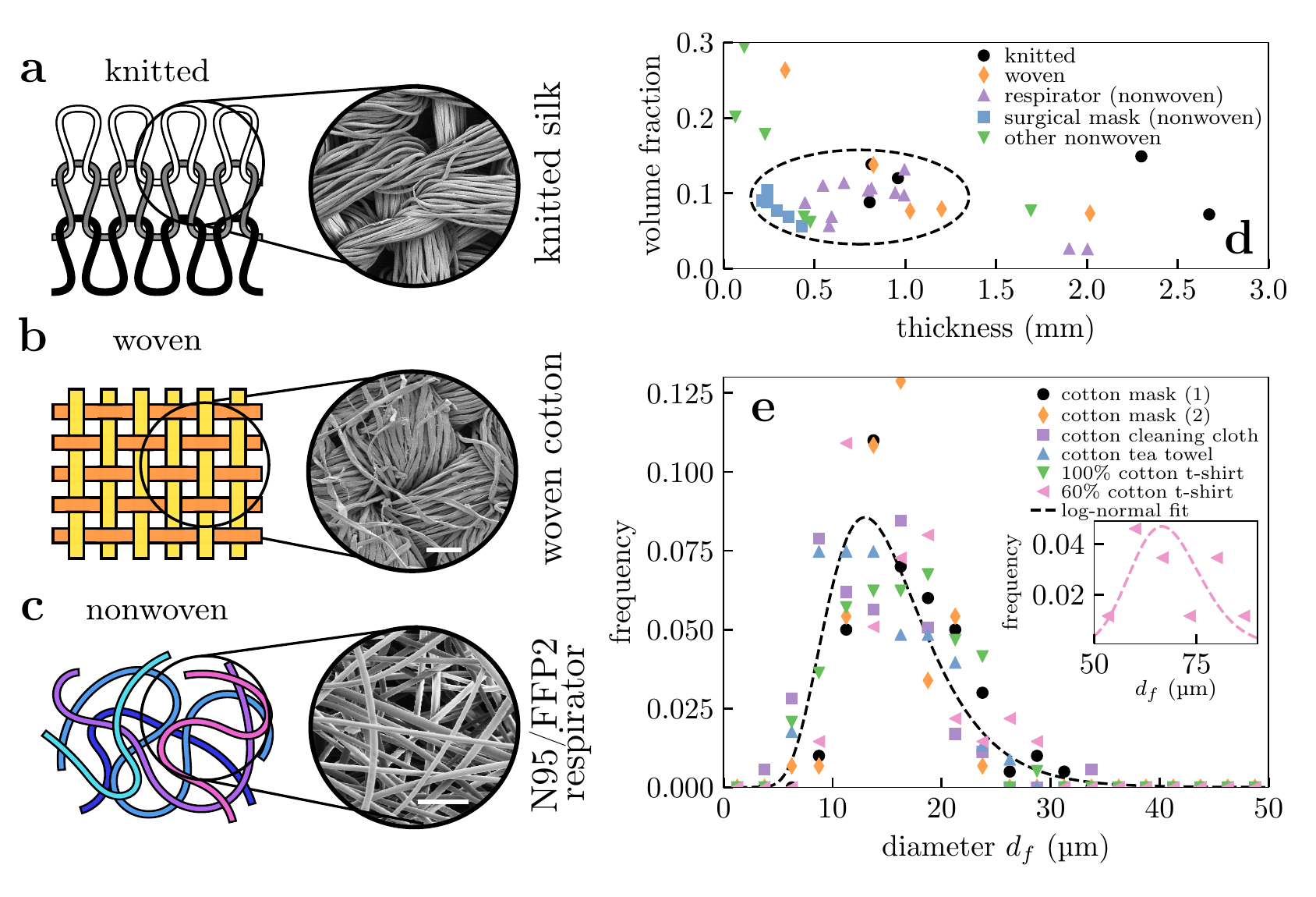}
  \caption{
    (colour online) Summary of fabrics comprising masks considered here.
    (a) Knitted fabrics formed by looping yarn through previous layers (layers coloured differently for clarity).
    (b) Woven fabrics formed by intersecting perpendicular yarns (the ``warp'' and ``weft'').
    (c) Nonwoven fabrics are formed by entangling fibres through other means, resulting in less ordered arrangements.
    Scanning electron microscope images of example fabrics in figures (a)-(c) share a scale bar of \SI{100}{\um}.
    (d) Geometric properties measured for sample fabric layers, with region of interest marked with a dashed circle (discussed in text).
    Respirators and surgical masks are comprised of multiple layers, with individual layers plotted separately within this panel.
    (e) Distribution of fibre diameters in cotton fabric samples, which loosely follow a log-normal distribution.
    Inset: the 60\% cotton 40\% polyester t-shirt shows a second peak at larger fibre diameter corresponding to the second material, which can also be modelled as a log-normal (pink dashed).
  }
  \label{fig:fabrics}
\end{figure*}

\section{Material properties of masks}
\label{sec:materials}

Fabrics are broadly categorised as \emph{knitted}, \emph{woven} or \emph{nonwoven}.
We refer to face coverings that would be worn by members of the public, that are neither surgical masks nor respirators, as cloth masks, and we use \emph{masks} as a catch-all term for all kinds of filters.
Filtration theory is well-developed for nonwoven materials \cite{wang2013}, which are typical of surgical masks and respirators.
However cloth masks typically contain knitted or woven fabrics so we introduce some fundamental characteristics of these fabrics below.

Knitted and woven fabrics are created by spinning fibres into yarn \cite{warren2018}.
In practice many of these threads are typically twisted together (the ``ply'') into a composite yarn with additional stability against being unwound.
Note that the process described above is for \emph{staple yarn}, where the natural fibres are short, but a different process (\emph{filament yarn}) may be used where the fibres are naturally long (\eg silk or synthetic polymers) which results in smoother thread (\cf silk strands are smooth in \Figref{fig:fabrics}(a) whereas cotton thread in \Figref{fig:fabrics}(b) features stray strands resembling a frayed rope).

Weaving involves interlacing multiple parallel yarn into a tight pattern, whereas knit fabrics are formed by drawing the yarn in complex loops (the ``stitches'').
Knitting thus results in regions of high curvature, so threads are able to bend which typically results in stretchier fabrics.
By contrast, nonwoven materials are formed by entangling the fibres mechanically, thermally or chemically which results in a less ordered structure.

The filtration characteristics of masks depends on many parameters, including the size and charge on the droplets as well as mask properties such as fibre thickness, density of fibres, their material composition and thickness of the mask.
In addition, in cloth fabrics details of yarn structure and weave/knit pattern matter.
Treating all of these within a single framework represents a significant challenge, so we focus on the most relevant parameters.

\subsection{Contact forces}

All combinations of fibres and droplets interact on contact between the droplet and the fibre, even when they are electrically neutral.
In almost all cases we expect droplets to stick when they contact the surface of the fibre.
Whether a droplet sticks and spreads on a surface it contacts, or carries on moving, is controlled by the ratios of two competing energies.
The first energy acts to keep droplets moving without sticking: the inertial or kinetic energy.
The second energy drives sticking and spreading of the droplets: the surface free energy.

For droplets in the size range of interest the surface free energy is much larger than the kinetic energy, so the surface free energy will win and the droplet will stick --- at least in the vast majority of cases.
The ratio of the kinetic energy to the surface free energy is the Weber number:
\begin{equation*}
  \mathrm{We} = \frac{\textrm{kinetic energy}}{\textrm{surface free energy}}=
  \frac{\rho_p d_p U_0^2}{\gamma},
\end{equation*}
for a droplet of mass density $\rho_p$, diameter $d_p$, surface tension $\gamma$, and moving at speed $U_0$.

For mucus droplets, $\gamma \sim \SI{0.05}{\newton\per\metre}$ \cite{gittings2015}.
For a droplet of diameter $d_p \sim \SI{10}{\um}$ travelling at \SI{0.1}{\metre\per\second}, $\mathrm{We}\sim \num{2e-3}$; surface tension forces are then about 500 times stronger than inertial forces, so we expect them to dominate and the vast majority of droplets to stick on contact.
Natural fibres such as cotton are more hydrophilic than synthetic polymers used in  medical-grade surgical masks and respirators.
However, at these very small Weber numbers we do not expect this variation to have a significant effect.
Small droplets can even stick to hydrophobic surfaces \cite{zwertvaegher2014}.

\subsection{Experiments}
\label{sec:experiments}

We examined a variety of fabrics used to make masks including cloth masks, surgical masks and respirators.
These masks are typically multi-layered structures, and were decomposed into their individual layers for examination.
Their properties are summarised in \Figref{fig:fabrics}(d) and a full breakdown is given in Table \ref{tab:fabrics} in the Supplementary Material (SM).

An important quantity for filtration is the volume fraction of fibres $\alpha$, which we determined from
\begin{equation}\label{eq:volume-fraction}
    \alpha = \frac{\rho_A}{\rho_b L},
\end{equation}
where $\rho_A$ is the areal density (the ``fabric weight'', typically measured in $\si{\gram\per\metre\squared}$), $\rho_b$ is the bulk density of the fibre and $L$ is the fabric thickness.
$\rho_A / L$ gives the fabric density.
We measured $\rho_A$ by weighing strips of known area, and $\rho_b$ is determined from the fabric material (\eg \SI{1.54}{\gram\per\metre\cubed} for cotton).
We measured the fabric thickness by cutting the material into thin strips, clamping them at one end, and measured their thicknesses under bright-field microscopy (Leica DMI 3000B) with a 4x and 10x objective (depending on the thickness of the fabric).
This method likely overestimates the thickness for fabrics with a yarn structure: an alternative method for \emph{inferring} the fabric thickness will be introduced in section \ref{sec:woven} (and a comparison of both methods is given in SM).
The manufacturers did not state the material composition of the surgical masks and respirators we sampled, so we assumed they were made from polypropylene fibres ($\rho_b = \SI{0.91}{\gram\per\metre\cubed}$).
We neglect any porosity \emph{within} the fibre in \eqref{eq:volume-fraction}; the SEM images in \Figref{fig:fabrics}(a-c) and the SM suggests that the porosity is not large enough to significantly affect the measured volume fractions.

We found that the majority of fabric layers were \SIrange{0.4}{1.2}{\milli\metre} thick consistent with \eg \Refcite{li2006} and had volume fractions in the range $0.05 \lesssim \alpha \lesssim 0.15$; these ranges are circled in \Figref{fig:fabrics}(d).
Notable exceptions to the latter rule included a silk tie and a paper bag with $\alpha \sim 0.26$ 0.20 respectively; however we found these samples to be difficult to breath through when placed to the face, making them unsuitable as potential mask materials.

For scanning electron microscopy (SEM) characterisation, samples were mounted on SEM stubs and coated with gold/palladium in an Emitech K575X Sputter coater before being imaged in an FEI Quanta 200 FEGSEM (Thermo Fisher Scientific).
SEM images were taken at \SI{8}{\kilo\volt} using comparable magnifications for all the fabrics.
From these images we manually measured the distribution of fibre diameters $d_f$, using the open-source software Fiji \cite{schindelin2012}, and parameterised it with a log-normal fit.
Natural fibres (\eg cotton) do not have perfectly circular cross-sections, so modelling them as cylinders is an approximation.
Our measured distribution of fibre diameters will thus be affected by fibre orientation, a consequence of obtaining 3d information from 2d images.
A minimum of 50 individual fibres were measured per fabric.
The size distributions obtained for cotton samples in \Figref{fig:fabrics}(e), and the remaining distributions are given in the SM.
For cotton we find $\ln{(d_f/\si{\um})} \sim \mathcal{N}(\mu=2.68, \sigma^2=0.12)$, so a cotton layer $\sim\SI{1}{\milli\metre}$ thick will typically be \numrange{50}{100} fibres thick.

\section{Capture of droplets by a single fibre}
\label{sec:single-fibres}

In this and the next section we describe the standard theory for filtration of droplets/particles, test its assumptions and generalise it to incorporate the polydisperse fibre size distributions obtained in the previous section.
In this section, we explore how a single fibre can collect droplets, and in the next section we look at filtration by a fabric formed from a mesh of such fibres.
We mostly follow \Refcite{wang2013}, but we also make use of \Refscite{chen1993,hinds1999}.
We use the subscript $f$ for fibre and $p$ for incident particles, \eg $d_p$ is the particle diameter whereas $d_f$ is the fibre diameter.

\subsection{Single-fibre efficiency from idealised flows}

\begin{figure}
  \centering
  \includegraphics[width=0.8\linewidth]{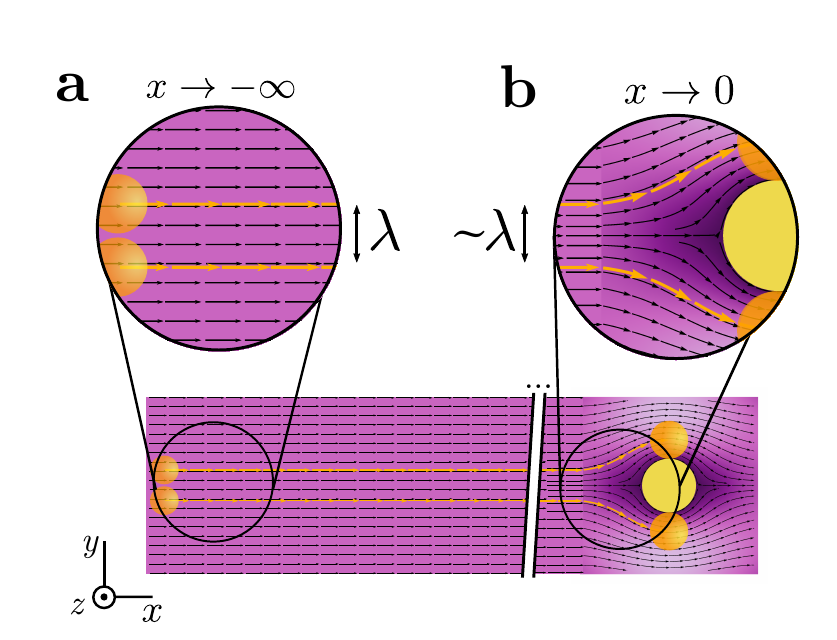}
  \caption{
    (colour online) Illustration of single-fibre filtration.
    Particles moving along trajectories between the upper and lower orange lines collide with the fibre and are filtered out.
    Particles along these trajectories just glance the surface of a fibre.
    The width of the collection window, $\lambda$ is defined as being the distance between the upper and lower trajectories far from the fibre, illustrated in (a).
    Far from the fibre we assume that particles follow the air streamlines.
    (b) Near the fibre, particle trajectories are highly curved precluding a simple geometric interpretation of $\lambda$.
    $\lambda$ depends on the particle and fibre sizes, as well as the background gas flow.
    Lighter (darker) shading corresponds to faster (slower) background flow speed.
    }
  \label{fig:lambda}
\end{figure}

To understand the filtering capacity of a single fibre, we consider the flow around an infinitely long cylinder aligned perpendicular to the direction of flow.
Assuming that the particles faithfully follow the streamlines infinitely far from the cylinder, we define the single-fibre efficiency as the fraction of particles collected by the fibre, \ie
\begin{equation}\label{eq:single-fibre-efficiency}
  \eta = \frac{\textrm{number of collection trajectories}}{\textrm{number of streamlines}}.
\end{equation}
Infinitely far from the mask the velocity field is $\vec{u} = U_0 \vec{e}_x$ so that the streamlines are distributed uniformly on planes with normal vector $\vec{e}_x$, as in \Figref{fig:lambda}(a).
We assumed $z$-symmetry so that our problem geometry is two-dimensional in the $xy$-plane, so this leaves width (in the $y$-direction) as a suitable measure for the number of streamlines.
Given these considerations we can write the single fibre efficiency as $\eta = \lambda/L_y$ where $\lambda$ is the width of the collection window in \Figref{fig:lambda} and $L_y$ is the total width of the mask in the $y$-direction.

Our definition of single-fibre efficiency differs from that normally used in filtration literature, namely the quantity $\lambda / d_f$ in \eg \Refscite{hinds1999,chen1993,wang2013}.
We have chosen a definition which guarantees $\eta < 1$ so it can be interpreted as a probability; the more common definition is \emph{not} properly normalised which can lead to incorrect and poorly posed results when combining multiple collection mechanisms (\cf section \ref{sec:combining-mechanisms}).

\begin{figure*}
  \includegraphics[width=\linewidth]{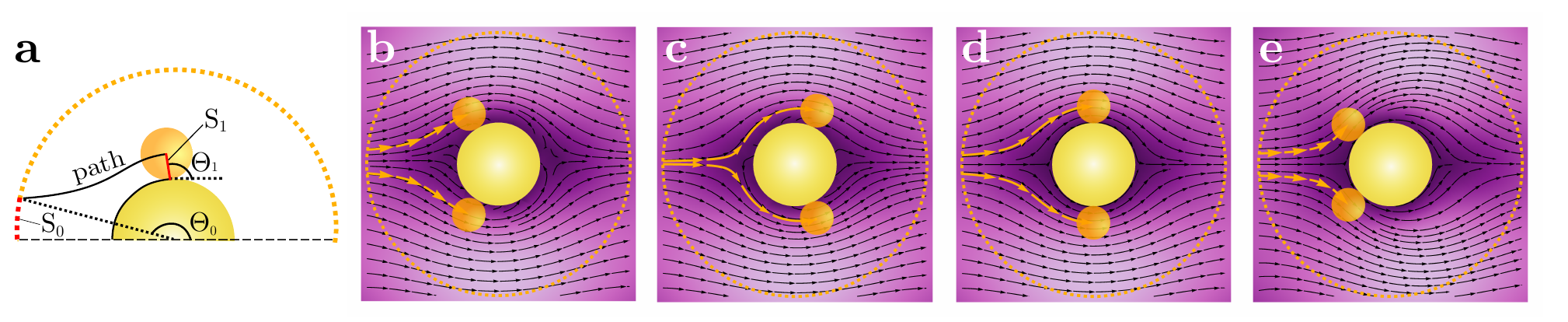}
  \caption{
    (colour online) Geometry of particle capture in the Kuwabara flow field.
    Lighter (darker) shading corresponds to faster (slower) flow speed.
    (a) Diagram of limiting trajectory: the particle path which only just collides with the fibre.
    In the absence of attractive forces and inertia the capture angle will be $\theta_1 = \pi/2$.
    (b-c) Effect of spherically symmetric forces on the incoming particle trajectories.
    The forces move the limiting trajectory towards the near or far side of the fibre depending on whether the interaction is attractive (b) or repulsive (c).
    (d-e) Inertia brings the limiting trajectories towards the near side of the collecting fibre, shown are particle trajectories for (d) $\St = 0$ and (e) $\St = 0.5$.
  }
  \label{fig:flow-field}
\end{figure*}

\subsubsection{Kuwabara flow field}
\label{sec:kuwabara}

Flow through a filter occurs at low Reynolds number, so it is well described by Stokes flow.
There is no solution to Stokes flow around a free cylinder because of the Stokes' paradox \cite{vandyke1975}, however the mask is composed of many fibres and we can obtain a solution for flow around a fibre immersed in an effective neighbourhood of other fibres: the \emph{Kuwabara flow} \cite{kuwabara1959}.
The effective neighbourhood is treated as an outer circle boundary at distance $a_f/\sqrt{\alpha}$ where $a_f$ is the radius of the fibre, so that the flow is modelled in the coaxial region $a_f \le \rho \le a_f/\sqrt{\alpha}$ which allows solution without a paradox.
Moreover, the radial component of the velocity at the outer boundary is taken as $u_\rho(\rho=a_f/\sqrt{\alpha}) = U_0 \cos{\theta}$.
$U_0$ is the average flow speed through the mask, obtained from the flow speed at the mask surface (\cf table \ref{tab:numbers}).

For incompressible flow $\nabla \cdot \vec{u} = 0$ the velocity field can be expressed in terms of a streamfunction, \ie
\begin{equation}\label{eq:u-streamfunction}
  \vec{u} = \vec\nabla \times \vec\psi
\end{equation}
where
\begin{subequations}
  \begin{align}
    \psi(\rho,\theta) =& U_0 f(\rho) \sin\theta \, \vec{e}_x, \\
    f(\rho) =& \frac{f_1}{\rho} + f_2 \rho + f_3 \rho^3 + f_4 \rho \ln{\left( \frac{\rho}{a_f} \right)},
    \label{eq:kuwabara-f}
  \end{align}
\end{subequations}
with coefficients $\{f_i\}$ set by the boundary conditions.
The Kuwabara flow field is obtained by assuming the velocity vanishes on the fibre surface $\vec{u}(\rho = a_f) = 0$, and that the vorticity $\vec\nabla \times \vec{u}$ vanishes at the outer boundary $\rho = a_f/\sqrt{\alpha}$ to approximate the neighbourhood around the fibre \cite{kuwabara1959}.
We give the explicit values of the coefficients obtained in the SM.

\subsubsection{Lattice Boltzmann flow field}

To test the validity of the Kuwabara flow field, we also calculated flow fields using Lattice Boltzmann (LB) simulations \cite{chenLB1998,zou1997,bao2011,behrend1994}.
In these simulations the Reynolds number $\Reynolds$ is nonzero, and can be varied, and the fluid is compressible.
However, at our small $\Reynolds$ the spatial variation in density is very small.
To do the LB simulations we use a modified version of a code from PALABOS group at the University of Geneva \cite{palabos2020}.
See SM for details.

We have performed two types of LB simulations.
In the first we can calculate the flow field around a single fibre, which allows us to calculate the single-fibre collection window $\lambda$.
In the second we calculate the flow field in a disordered hexagonal lattice of fibres, which is our model of a mask.
This flow field allows us to test the theory's ability to predict filtration efficiency, at least within our simple two-dimensional model.
In all cases we run the LB simulations until we reach steady state, and then use the steady-state flow field.

\subsubsection{Particle motion}

The equation for particle velocity $\vec{v}$ (Newton's second law) while being transported by the flow $\vec{u}$ is
\begin{equation}\label{eq:particle-newton2}
  m_p \frac{d \vec{v}}{d t} = - \frac{\vec{v} - \vec{u}}{B} + \vec{F}
\end{equation}
where $m_p$ is its mass.
The first term on the right hand side is the Stokes drag.
In this term $B = C/6\pi \mu a_p$ is the particle mobility, with $\mu$ the dynamic viscosity of air and $C$ the Cunningham slip correction factor \cite{lee1980,kanaoka1987}.
$\vec{F}$ contains any other external forces such as gravity, which we do not consider here.
We have assumed that the particle interacts with the flow field as a point particle so that: (a) the flow field $\vec{u}$ is unperturbed by the presence of the particle and (b) the Stokes drag couples only to the particle's centre of mass.

We denote dimensionless parameters with tildes, defined through the transformations $\vec{u} = U_0 \widetilde{\vec{u}}$, $\vec{v} = U_0 \widetilde{\vec{v}}$, $\vec{r} = a_f \widetilde{\vec{r}}$, and $t = a_f \widetilde{t} / U_0$ so \eqref{eq:particle-newton2} becomes%
\footnote{Note that we must use the fibre rather than particle size because $a_f$ is the only relevant lengthscale entering \eqref{eq:particle-newton2} as we have assumed that particles couple to the flow field as point particles.}
\begin{equation}\label{eq:stokes-newton2}
  \mathrm{St} \frac{d \widetilde{\vec{v}}}{d \widetilde{t}}
  =
  -(\widetilde{\vec{v}} - \widetilde{\vec{u}}) + \frac{B}{U_0} \vec{F},
\end{equation}
with Stokes number
\begin{equation}\label{eq:stokes-number}
  \mathrm{St}
  = \frac{m_p U_0 B}{a_f}
  = \frac{2 \rho_p a_p^2 U_0 C}{9\mu a_f}
  \sim \frac{\num{6.2e6}}{\si{\metre\squared\per\second}} \frac{d_p^2}{d_f} U_0 C,
\end{equation}
with the latter step evaluated for parameter values typical of incoming droplets.
These are in table \ref{tab:numbers}.
The Stokes number describes the effective inertia of the particle moving under the flow field.
For threads with diameter $\mathcal{O}(\SI{100}{\um})$ typical of yarns used in knitted and woven fabrics, we find $\St \ll 1$ making inertia irrelevant for particles around  $\mathcal{O}(\SI{1}{\um})$ in diameter; for this reason the smaller fibres are crucial for capture of exhaled droplets in cloth masks.

\begin{table}[b!]
  \begin{center}
  \begin{ruledtabular}
  \begin{tabular}{ccc}
    Quantity & Value & Reference \\
    \hline
    \multicolumn{3}{c}{
    \textbf{Air}} \\
    \hline
    mass density & \SI{1.2}{\kilogram\per\metre\cubed} & \onlinecite{crc} \\
    dynamic viscosity $\mu$ & \SI{1.8e-5}{\pascal\second} & \onlinecite{crc} \\
    kinematic viscosity $\nu$ & \SI{1.5e-5}{\metre\squared\per\second} & \onlinecite{crc} \\
    \hline
    \multicolumn{3}{c}{
    \textbf{Water/mucus}} \\
    \hline
    mass density $\rho_p$ (water) & \SI{998}{\kilogram\per\metre} & \onlinecite{crc}\\
    dynamic viscosity (mucus) & \SI{0.1}{\pascal\second} & \onlinecite{gittings2015} \\
    mucus/air surface tension
    $\gamma$  & $\SI{0.05}{\newton\per\metre}$ & \onlinecite{gittings2015} \\
    \hline
    \multicolumn{3}{c}{\textbf{Typical breathing flow rates}} \\
    \hline
    tidal breathing at rest
    & \SI{6}{\litre\per\minute} & \onlinecite{caretti2004} \\
    during mild exertion
    & \SI{20}{\litre\per\minute} & \onlinecite{caretti2004} \\
    during moderate exertion
    & \SI{30}{\litre\per\minute} & \onlinecite{caretti2004} \\
    during maximal exertion
    & \SI{85}{\litre\per\minute} & \onlinecite{caretti2004} \\
    \hline
    \multicolumn{3}{c}{\textbf{Average flow speeds}} \\
    \hline
    effective mask area & \SI{190}{\centi\metre\squared} & \onlinecite{coffey1999} \\
    flow speed (rest)
    & \SI{0.5}{\centi\metre\per\second} \\
    flow speed (mild)
    & \SI{1.8}{\centi\metre\per\second} \\
    flow speed (moderate)
    & \SI{2.7}{\centi\metre\per\second} \\
    flow speed (maximal)
    & \SI{7.5}{\centi\metre\per\second} \\
  \end{tabular}
  \end{ruledtabular}
  \caption{
    Table of key parameter values for masks including air, water and mucus at \SI{20}{\celsius} and atmospheric pressure \SI{e5}{\pascal}.
    Note that small droplets dry rapidly and this will cause their viscosity to increase.
    Flow rates are determined from the volume typically exhaled during one minute.
    Moderate exertion is defined as that readily able to be sustained daily during 8 hours of work, whereas maximal exertion is the upper limit of what can be sustained for short periods of time (\eg during competitive sports).
    Flow speeds are calculated for the stated mask area and flow rates assuming perfect face seal; in practice leakage would reduce flow through the mask.
  }
  \label{tab:numbers}
  \end{center}
\end{table}

\subsubsection{Particle deposition and collection efficiency}

For the LB flow field the length of the single-fibre collection window $\lambda$ can be determined by direct measurement of its geometric definition in \Figref{fig:lambda}.
The Kuwabara flow field is only valid in the region of high curvature close to the fibre surface, so determining $\lambda$ is slightly more subtle.

Defining $n$ as the number density of incoming particles, the continuity equation in the steady-state $\dot{n} = 0$ yields $\vec\nabla \cdot (n \vec{v}) = 0$.
All particle trajectories that terminate on the fibre surface are contained in the volume bounded by the limiting path shown by a solid black line in \Figref{fig:flow-field}(a).
We integrate the continuity equation over this and apply the divergence theorem to give
\begin{equation}\label{eq:limiting-flux}
  \int_{S_0} n \vec{v} \cdot d\vec{S} + \int_{S_1} n \vec{v} \cdot d\vec{S} = 0
\end{equation}
using the fact that the $\vec{v} \cdot d\vec{S} = 0$ along the limiting trajectory and the fibre surface at $r=a_f$, and the surfaces $S_{\{0,1\}}$ are defined in \Figref{fig:flow-field}(a).
We write the magnitude of either integral in the above expression as $\Phi/2$: (half) the rate of particle deposition on the fibre surface.
We multiply by two to account for collection along both sides of the fibre, taking advantage of the symmetry in the $y$-direction.

The width of the collection window is determined from the deposition rate by $\lambda = \Phi/n_0 U_0 L_z$ where $n_0$ is the particle number density far away from the fibre and $U_0$ is the flow speed.
We apply the boundary condition $n(r=a_f/\sqrt{\alpha}) = n_0$, which is a constant along $S_0$, so we have the following expression for collection efficiency:
\begin{equation}\label{eq:lambda-limiting-trajectory}
  \lambda = \frac{d_f}{\sqrt{\alpha}} \int_\pi^{\theta_0}
  \widetilde{v}_\rho\left(\theta; \rho=\frac{a_f}{\sqrt{\alpha}}\right) \, d\theta.
\end{equation}
The velocity field at the outer boundary is a boundary condition of the field, so $\theta_0$ is the key quantity needed to evaluate efficiency through this route.
For $\vec{v} = \vec{u}$ at the boundary this reduces to
\begin{equation*}
  \lambda = d_f \sin{(\theta_0)} f\left(\frac{a_f}{\sqrt{\alpha}} \right).
\end{equation*}
The angle $\theta_0$ is obtained by following the limiting trajectory (\eg the one shown in \Figref{fig:flow-field}(a)) that only just glances the fibre.
Particle trajectories in this limit are defined by
\begin{equation}
  \frac{1}{\rho}\frac{d\rho}{d\theta} = \frac{v_\rho}{v_\theta} = \frac{u_\rho}{u_\theta},
\end{equation}
which can be integrated backwards in time with final conditions $r = a_f$ and $\theta = \theta_1 = \pi/2$ to determine $\theta_0$.

\subsubsection{Single-fibre efficiency from combined mechanisms}
\label{sec:combining-mechanisms}

From the definition of the single-fibre collection efficiency \eqref{eq:single-fibre-efficiency}, we can see that if the mechanisms act completely independently then the \emph{penetration probability}, the probability of passing the fibre, will be the product of the penetration probabilities due to the individual mechanisms \ie
\begin{equation*}
    1 - \eta
    =
    \prod_k \left( 1 - \frac{\lambda_k}{L_y} \right)
    =
    1 - \sum_k \frac{\lambda_k}{L_y}
    + \mathcal{O}\left( \left(\frac{\lambda_k}{L_y}\right)^2 \right)
\end{equation*}
where $k$ sums over the different mechanisms and the last step is valid in the macroscopic limit $(\lambda/L_y)^2 \ll 1$.
However, in practice these mechanisms are not independent and the relative catchment lengths $\lambda_k$ will overlap.
Assuming perfect overlap and no interaction between mechanisms, the total efficiency will simply equal the most efficient individual mechanism \ie $\max{(\{\eta_k\})}$.

Combining the two limits above, we find
\begin{equation*}
  \frac{\max{(\{\lambda_k\})}}{L_y}
  \le
  \eta
  \le
  \sum_k \frac{\lambda_k}{L_y}
\end{equation*}
If one mechanism dominates over the others then these two bounds converge and we can simply take the dominant mechanism.

\subsubsection{Specific mechanisms}

As noted in the introduction, there are four principle mechanisms by which droplets may be captured by a mask which concern us here,
\emph{steric interception, inertial impaction, diffusion} and \emph{electrostatic capture} \cite{wang2013}.
These mechanisms generally act in different size regimes, so it is simpler to calculate their effects in isolation and then combine them using the approach outlined in the previous section.
The SARS-CoV-2 virus is $\sim \SI{0.1}{\um}$ in diameter, so this is the smallest size of interest.
Exhaled droplets have been observed across the wide range of $\mathcal{O}(\textrm{\LineSIrange{0.1}{100}{\um}})$, which corresponds to Stokes numbers from \numrange{e-4}{e3}.
However, the majority of droplets are larger than $\ge \SI{1}{\um}$ \cite{johnson2011,gregson2021} where $\St \gtrsim \textrm{\Linenumrange{e-3}{e-2}}$, and coarser droplets are expected to contain more virus on average \cite{freitag2020,robinsonPanic2020}.
The $\ge \SI{1}{\um}$ size regime is therefore of most interest, and the importance of the finer regime $\mathcal{O}(\textrm{\LineSIrange{0.1}{1}{\um}})$ will be scenario-dependent.

Electrostatic capture is crucial for high efficiency filtration of particles with size of order $\mathcal{O}(\SI{0.1}{\um})$ in respirators which make use of electret fibres that sustain surface charges $\sigma_0$ of order $\mathcal{O}(\SI{1}{\nano\coulomb\per\centi\metre\squared})$ \cite{chen1993,kravtsov2000}
\footnote{Note that the dielectric breakdown of air would occur for cylinders with a surface charge density in the range of $\sim$\SIrange{3}{10}{\nano\coulomb\per\centi\metre\squared} (depending on the fibre's dielectric constant \cf SM for this calculation), so electret fibres are impressively capable of sustaining almost the maximum possible charge.}.
The electrostatic forces in electrets are typically an order of magnitude more efficient at capture than the mechanical forces, and this efficiency is expected to scale as $\propto \sigma_0$ for the Coulombic force or $\propto \sigma_0^2$ for the dielectrophoretic force \cite{chen1993}.
However, the surface charge density is typically two orders of magnitude smaller in cloth masks
\footnote{Natural cellulose fabrics such as cotton and wool can typically sustain a maximum charge density in the range of $\mathcal{O}(\SI{0.01}{\nano\coulomb\per\centi\metre\squared})$ (or $\mathcal{O}(\SI{0.1}{\nano\coulomb\per\centi\metre\squared})$ for silk) when charged tribolectrically \cite{liu2018}.
This provides an upper bound for charge in most cloth fabrics, and we expect this to have minimal impact on filter efficiency.
Electrostatics could become important for other fibres made from synthetic polymers such as polyester or polypropylene that can sustain more charge \cite{liu2018}, but for most fabrics it can be neglected.}
and so electrostatic capture should be an order of magnitude less efficient than for the first three mechanical mechanisms.
We therefore neglect electrostatic capture in this work.

For interception, collection occurs when the finite-sized particles touch the surface of the fibre while passing, with the limiting trajectory occurring at $\theta_1=\pi/2$.
The particle follows the flow $\vec{v} = \vec{u}$ (inertia is included in \emph{impaction} but not in interception) and the limiting trajectory occurs at $\theta_1=\pi/2$, so \eqref{eq:lambda-limiting-trajectory} gives
\begin{equation}\label{eq:interception-lambda}
  \lambda_R = \frac{2\psi(a_f + a_p, \pi/2)}{U_0}.
\end{equation}
In general, capture efficiency is further enhanced by diffusion and inertia.
The role of diffusion is quantified by the \emph{P\'{e}clet number},
\begin{equation}\label{eq:peclet}
  \Pe = \frac{\textrm{rate of convection}}{\textrm{rate of diffusion}}
  = \frac{d_f U_0}{D},
\end{equation}
where $D$ is the particle diffusion coefficient for motion relative to the flow.
We find that $\Pe \ll 1$ for $d_p \gtrsim \SI{1}{\um}$ so diffusion is negligible for capture of larger droplets.
Similarly, inertia plays no role in the capture of smaller droplets $d_p \lesssim \SI{0.1}{\um}$ because $\St \ll 1$ in that regime.
Most exhaled droplets are larger than $d_p \gtrsim \SI{1}{\um}$ \cite{johnson2011,gregson2021}, thus inertia is crucial to the effectiveness of cloth masks in the relevant size regime and warrants a more detailed treatment.
We use standard results for diffusion, given in the SM.

To determine the single-fibre collection window $\lambda$ for finite Stokes number $\St$, we use an iterative scheme where we test whether a particular initial angle leads to collision with the fibre, and update a lower and upper bound for $\theta_0$ accordingly.
By testing for collision for the midpoint between the current bounds, we ensure each iteration adds $\sim 1$ bit of information to the approximation of $\lambda$ and convergence is rapid.
For the LB flow field we use a similar scheme, but varying the initial height of the particle far from the mask where the flow is parallel (\cf \Figref{fig:lambda}).

\begin{figure}
\includegraphics[width=0.9\linewidth]{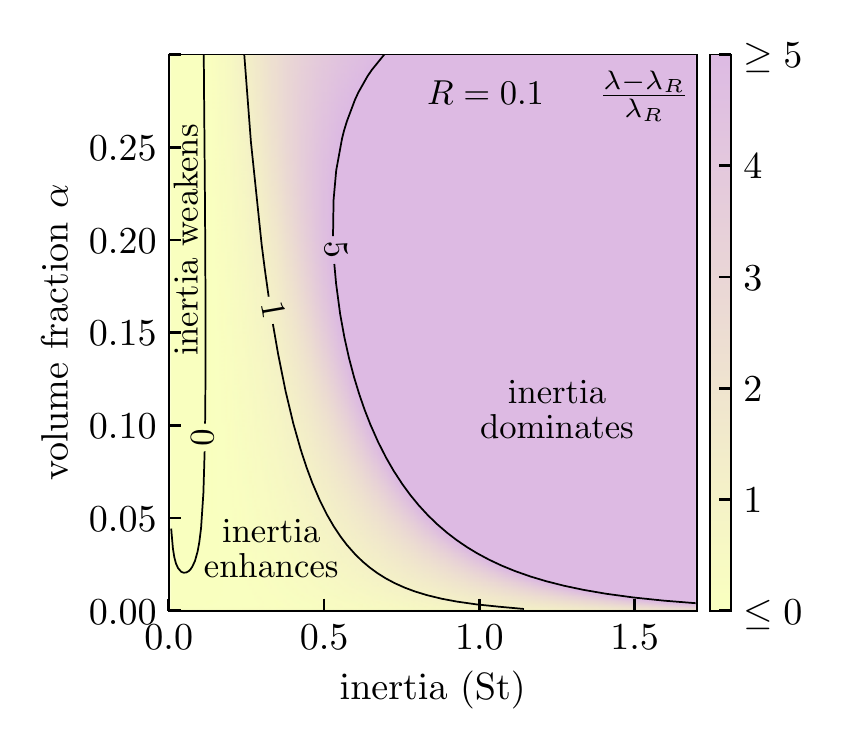}
  \caption{
    (colour online)
    Deviation $\lambda/\lambda_R - 1$ of single-fibre collection efficiency $\lambda$ from the interception capture efficiency $\lambda_R$ for finite particle-to-fibre size ratio $R = d_p / d_f = 0.1$.
    We see a sharp crossover from interception to inertial capture as the dominant mechanism.
    $\lambda$ increases by a factor of $\gtrsim 5$ as $\St$ is increased to $\sim$0.5.
    $\lambda_R$ is defined in \eqref{eq:interception-lambda}.
    We assumed the particle moves in the Kuwabara flow field in these calculations.
  }
  \label{fig:phase-diagram}
\end{figure}

\subsection{Droplet inertia rapidly increases efficiency above a threshold value}

Inertia causes droplets to deviate from streamlines which can bring particles closer to the fibre enhancing capture.
The inertia, as measured by the Stokes' number $\St$ in \eqref{eq:stokes-number}, increases as $d_p^2$ so this mode dominates capture of large droplets.
Naively, we would expect this increase in efficiency to be a simple increasing function of the Stokes number.
However, inertia also carries particles closer to the fibre where the flow is slower and more curved, which increases the opposing forces acting against the particle; this creates competition and inertial capture is non-trivial for intermediate values of $\St$.

In \Figref{fig:phase-diagram} we show how $\lambda$ varies with $\alpha$ and $\St$.
There is a sharp crossover from weak to strong capture as $\St$ reaches values in the $\mathcal{O}(0.1)$ range when $\alpha \gtrsim 0.01$.
This sharp crossover is a residual of an underlying dynamical transition occurring in the point particle limit $d_p/d_f \to 0$ demonstrated by \citeauthor{araujo2006}~\cite{araujo2006}.
We will explore this transition in more detail in a future manuscript, but here the important message is that once inertia becomes a relevant mechanism the total mask efficiency will rapidly increase (with particle size) to unity independent of the mask details.
However, the location of this crossover does depend on the mask properties.
Curiously, we find that for small $\St$ there is a region where inertia \emph{decreases} the efficiency of capture for finite $R$ highlighting that capture efficiency has a non-trivial dependence on inertia.

\begin{figure}
  \includegraphics[width=0.9\linewidth]{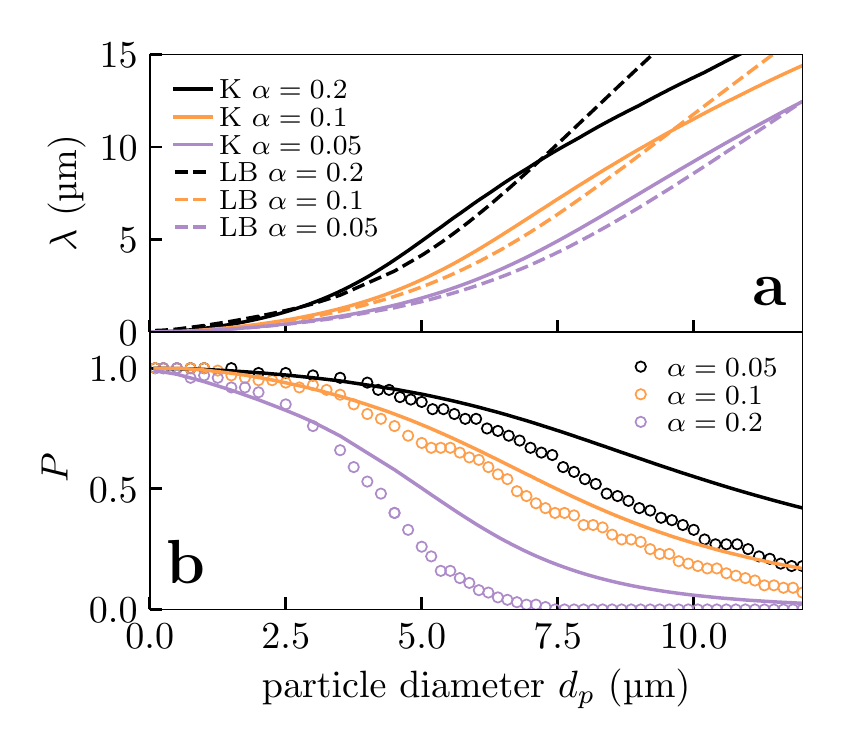}

  \caption{
    (colour online) Comparison of theoretical model against Lattice Boltzmann simulations.
    (a) Plot of the single fibre $\lambda$ as a function of particle diameter calculated from the Kuwabara (solid lines) and LB (dashed lines) flow fields.
    (b) Comparison between the penetration $P$ calculated using LB simulations of model filters (points) with the predictions of \eqref{eq:penetration-length} (curves).
    In both cases, the flow speed $U_0 = \SI{2.7}{\centi\metre\per\second}$ and the fibre diameter $d_f=\SI{15}{\micro\metre}$ with $\alpha=0.05$, $0.1$ and $0.2$.
  }
  \label{fig:LBtest}
\end{figure}

All the above calculations used the approximate Kuwabara flow field to compute $\lambda$.
We performed LB simulations to check the validity of the Kuwabara approximation.
Kuwabara and LB values for $\lambda$ are compared in \Figref{fig:LBtest}(a).
We note that, especially at small fibre volume fraction $\alpha$, the Kuwabara approximation gives $\lambda$ values close to those obtained by LB simulations.
So we conclude that at least under most conditions, the Kuwabara flow field yields good approximations for $\lambda$.

Above the dynamical transition, $\lambda$ increases rapidly with particle size, see \Figref{fig:LBtest}(a), due to the effect of increasing inertia.
So in this regime, typically of particles micrometres in diameter, the filtration efficiency increases rapidly.
To see this, consider a fibre of diameter $\SI{15}{\um}$ (typical of cotton from \Figref{fig:fabrics}(e)), in air for a flow speed of $\SI{2.7}{\centi\metre\per\second}$ corresponding to breathing during moderate exertion.
LB calculations for a particle of diameter $\SI{2}{\um}$ find a collection range $\lambda=\SI{0.36}{\um}$ or about 2.5\% of the fibre width.
However, increasing the particle diameter to $\SI{8}{\um}$ yields a collection range $\lambda=\SI{7.1}{\um}$ or almost half the fibre width.

\section{From single fibres to total filter efficiency}
\label{sec:total-filter}

In the previous section we developed the theory for the width of the region over which a single fibre collects the droplets: $\lambda$.
In this section we model a filter as an array of these fibres, and calculate filtration efficiencies from $\lambda$, the volume fraction $\alpha$ and thickness of the filter.
Standard filtration theory assumes the fibres are identical in shape and size, act (\ie filter) independently and are distributed homogeneously in space.
These assumptions are reasonable for nonwoven materials such as surgical masks,
however in common fabrics we typically find:
\begin{enumerate}
\item The individual fibres vary in shape and size.
\item In woven and knitted fabrics, the fibres are hierarchically arranged because of the yarn structure.
  The fibres are densely packed in yarns, leaving regions of lower density in the inter-yarn pores.
\end{enumerate}
Our treatment generalises filtration theory to account for these heterogeneities.
We present these generalisations in subsequent sections, and numerically compare the resulting theory against the experimental data available from the literature.

\subsection{Filter efficiency from a polydisperse assembly of fibres}
\label{sec:polydisperse-filtration}

Standard filtration theory considers filters as an assembly of identical cylindrical fibres.
Here, we borrow ideas from statistical mechanics to rigorously formulate the main result of filtration theory, as well as provide the natural generalisation for when the fibres vary in diameter.
As we noted in section \ref{sec:experiments}, natural fibres are seldom perfectly cylindrical so this formulation is approximate.

For simplicity we consider a rectangular filter of dimensions $(L_x, L_y, L_z)$, although the shape details perpendicular to the direction of flow do not matter because we will ultimately consider the limit of an infinite plane.
On average the streamlines (carrying particles) will occupy an effective area of $(1-\alpha) A$, so the effective efficiency is modified to $\eta_k = \lambda_k/((1-\alpha) L_y)$, where we have introduced a subscript $k$ for the efficiency of fibre $k$ as materials are generally heterogeneous and $\lambda$ will be taken from a distribution of values (\cf distribution of fibre sizes in \Figref{fig:fabrics}(e)).
Assuming the results for single fibres of previous sections, the probability that a particle is collected by fibre $k$ then equals the probability that a cylinder of diameter $\lambda_k$ crosses the particle path.
Those results assume that all the fibres are aligned perpendicular to the flow direction.

In the simplest case where the particle trajectory is a straight line through the filter, the probability that a particle passes the $k$\textsuperscript{th} fibre is $P^{(1)}_k = 1 - \eta_k$.
Assuming the fibres act independently gives the \emph{penetration}, the total fraction of particles that pass through the filter, as
\begin{equation*}\label{eq:multiplicative-penetration}
  P = \lim_{L_y \to \infty} \prod_{k=1}^N P^{(1)}_k
\end{equation*}
where $N = n L_x L_y$ is the total number of fibres in terms of fibre density (number per unit cross-sectional area) $n = 4\alpha/\pi d_f^2$.
Geometrically, the $L_y \to \infty$ limit above takes the limiting geometry as an infinite plate (as $L_z \to \infty$ is already implicit in our 2d formulation).
We take this limit by considering the logarithm of both sides, giving
\begin{equation*}
    \ln{P}
    =
    \lim_{L_y \to \infty} n L_x L_y \int_{\mathbb{R}^+} \ln{p(d_f)} \, d\mu(d_f)
\end{equation*}
which introduces the measure on the fibre size distribution $\mu(d_f)$ that is normalised through $\int_{\mathbb{R}^+} d\mu(d_f) = 1$.
Taking the limit yields
\begin{equation*}
  \lim_{L_y \to \infty} L_y \ln{\left( 1 - \frac{\lambda}{(1-\alpha) L_y} \right)}
  =
  - \frac{\lambda}{1-\alpha},
\end{equation*}
so the total penetration becomes
\begin{subequations}\label{eq:total-penetration}
  \begin{align}
    P
    =&
    \exp{\left( - \frac{L_x}{\xi} \right)},
    \intertext{with penetration length}
    \label{eq:penetration-length}
    \xi
    =&
    \frac{(1 - \alpha) \pi}{4\alpha \overline{\lambda}}
    \int_{\mathbb{R}^+} d_f^2 \, d\mu(d_f),
    \intertext{and effective collection window}
    \overline{\lambda}
    =&
    \int_{\mathbb{R}^+} \lambda(d_f) \, d\mu(d_f).
  \end{align}
\end{subequations}
Finally, we take the measure to be a log-normal distribution based on the fits to the experimental measurements described in section \ref{sec:experiments} (\cf table \ref{tab:fabrics} in SM).

Our fundamental assumptions to achieve the above expressions were that (a) the fibres act independently, and (b) their sizes are independent and identically distributed random variables.
We directly test assumption (a) in section \ref{sec:fibre-independence} of the SM.

In \Figref{fig:LBtest}(b) we compare the predictions of \eqref{eq:total-penetration} with the penetrations observed in LB simulations of a disordered lattice of fibres.
We see that \eqref{eq:total-penetration} systematically overpredicts the penetration, but that the error is typically relatively small.
Thus, as the model is only a very simplified realisation of a mask, we conclude that the approximations involved in \eqref{eq:total-penetration} give an acceptable level of accuracy.
Note that due to the Stokes' paradox \cite{vandyke1975}, fibres are never completely independent of each other.
Moreover, fibres will be arranged in a disordered fashion and so there will be variation in the distances between neighbouring fibres, so \eqref{eq:total-penetration} essentially both neglects correlations and assumes each fibre has the same local environment.

\subsection{Filtration efficiency of nonwoven materials}

\begin{figure}
  \includegraphics[width=0.925\linewidth]{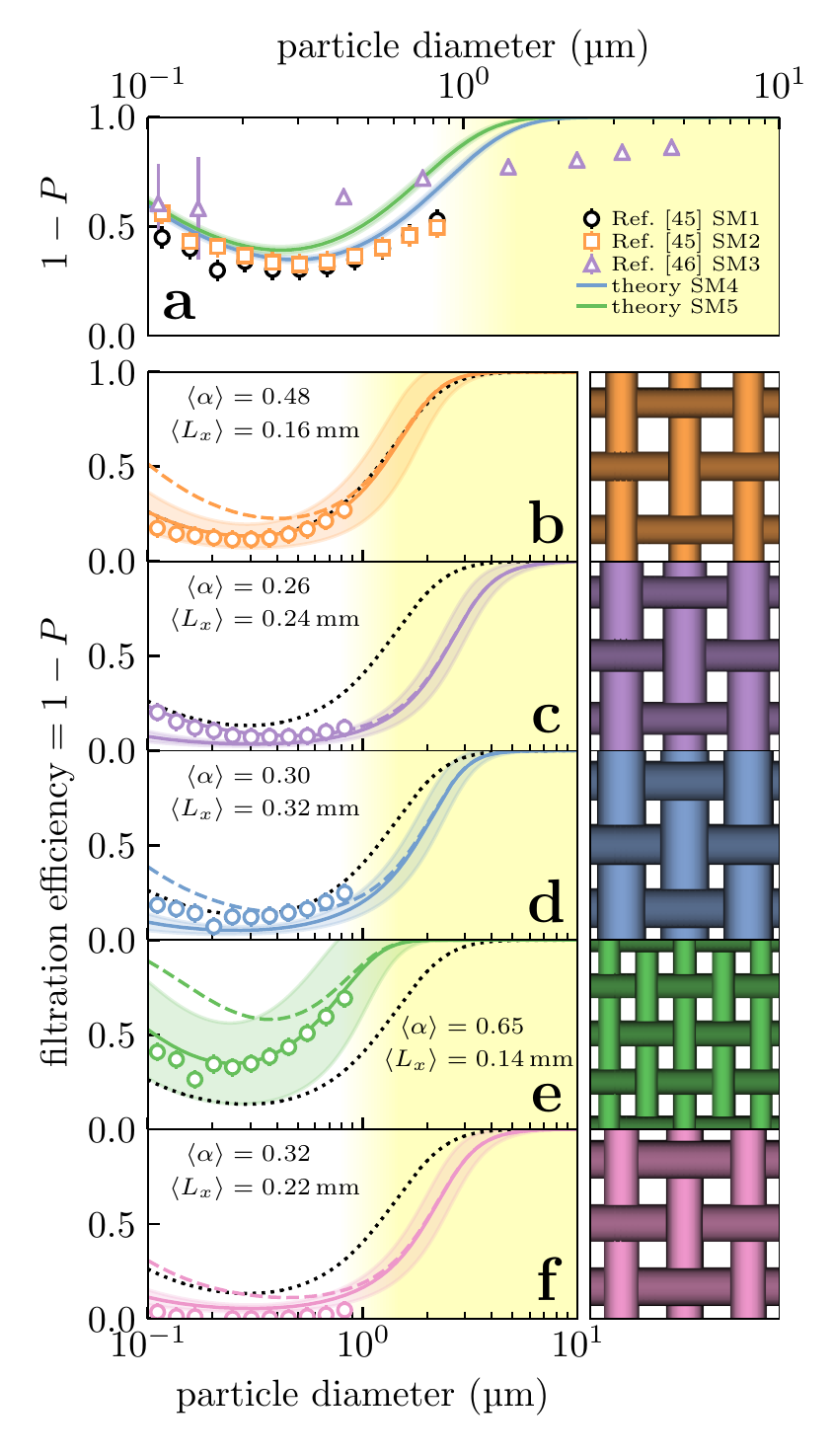}
  \caption{
    (colour online) Comparison between our theoretical model (lines) and the experimentally determined filtration efficiencies (points) of \Refscite{zangmeister2020,konda2020} for (a) surgical masks and (b-f) the plain-woven cotton fabrics considered in \Refcite{zangmeister2020} (numbered 1-4 and 11 there and in table \ref{tab:zangmeister-cotton}).
    The filled region surrounding the theoretical prediction indicates the confidence interval from propagating the uncertainties in the experimentally determined parameters.
    For reference, the left panels in (b-f) show our ``zeroth-order'' prediction where we ignore the inter-yarn pores (dashed) and 1/3 of surgical mask SM4 (black dotted).
    The right panels in (b-f) are illustrations of \SI{1}{\milli\metre^2} square regions of each fabric.
  }
  \label{fig:literature-compare}
\end{figure}

The theory of the previous section is sufficient to predict the filtration efficiency of nonwoven materials.
To demonstrate this we compare the predictions of our model against experimental data for three surgical masks from \Refscite{zangmeister2020,konda2020} (SM1, SM2 and SM3).
The physical properties of these masks were not stated, so for comparison we sampled two new surgical masks (SM4 and SM5) and characterised their thickness and fibre distribution using the methods in section \ref{sec:experiments}.
These surgical masks consisted of three layers with distinct properties and thus penetrations through individual application of \eqref{eq:total-penetration}.
Eq.\ \eqref{eq:total-penetration} implies that layers act independently, so the total mask penetration was obtained by combining the penetrations of the individual layers multiplicatively.

Our results compare favourably against the literature data in \Figref{fig:literature-compare}(a).
Our theoretical prediction for these masks closely matches the precise data of \Refcite{zangmeister2020} for their own masks (SM1 and SM2).
Our theory captures the experimental behaviour without any free parameters.
Moreover, our model agrees with the trend of increasing filtration efficiency going into the micron regime seen in \Refcite{konda2020} (SM3).
There was a small amount of variation in the physical properties we observed in masks SM4 and SM5 (parameters given in SM) which creates some variation in filtration efficiency.
The small deviations from the precise data of \Refcite{zangmeister2020} may therefore arise from differences in mask manufacture.

\subsection{Ease of breathing through a mask and the effect of hierarchical structure on the flow}
\label{sec:pressure-drop}

\begin{figure}
  \centering
  \includegraphics[width=0.9\linewidth]{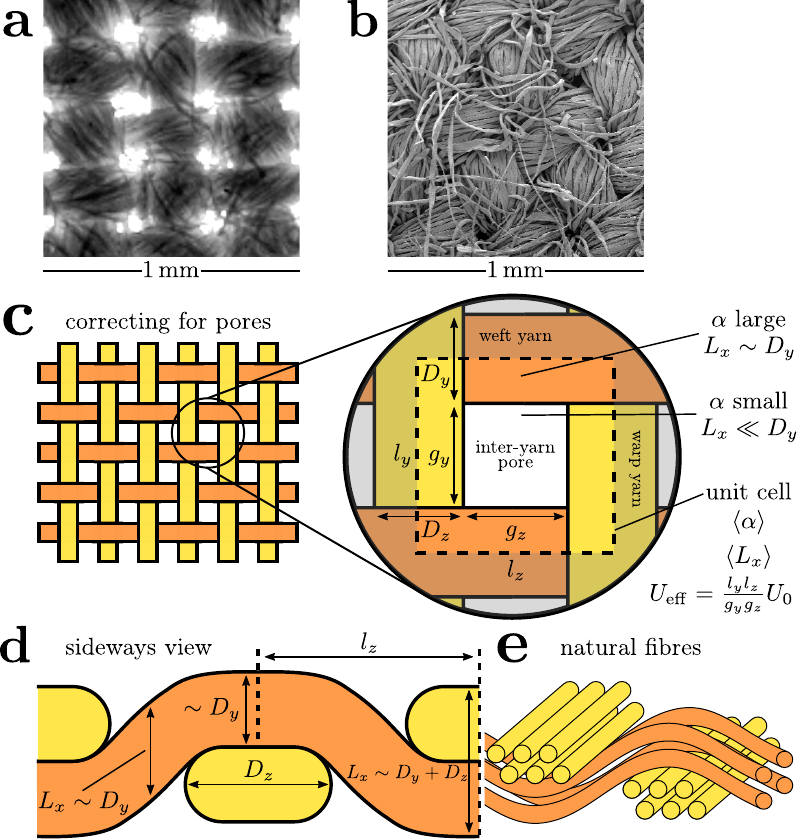}
  \caption{
    (colour online)
    (a-b) The same woven cotton layer under (a) optical and (b) scanning electron microscopy.
    (c) Schematic of how we treat heterogeneous woven fabric as an effective homogeneous medium by averaging over the geometric parameters over the dense yarn and sparse pore regions.
    (d) Sideways view of a yarn showing the local fabric thicknesses taken for averaging.
    Elastic deformations flatten the yarns' cross-sections into stadium shapes \cite{kemp1958}.
    (e) Idealised decomposition of yarns into their constituent fibres.
  }
  \label{fig:schematic}
\end{figure}

The pressure drop across a homogeneous filter $\Delta p$ is given by \cite{wang2013}
\begin{equation}
  \Delta p = \frac{\mu L_x U_0 f_p(\alpha)}{d_f^2},
  \label{eq:deltap}
\end{equation}
where the function $f_p(\alpha) = 16\alpha / K$ for the Kuwabara flow field or it can be estimated from previous empirical studies \cite{wang2013}.
The pressure drop across the mask needed for a given flow speed $U_0$, scales with this speed as well as mask thickness placing limits on how thick masks can be made.
The variation with fibre size as $d_f^{-2}$ (which follows directly from Poiseuille flow) makes finer fibres harder to breathe through.
This is often expressed in terms of a filter quality factor $q$ such that $P = e^{-q \Delta p}$ \cite{chen1993,wang2013}.

Pressure drops measured across masks vary from a few \si{\pascal} \cite{konda2020} to \SI{100}{\pascal} and above \cite{wangData2020}.
This pressure drop cannot be too large, to allow easy breathing.
The N95 standard specifies maximum values for $\Delta p$ of \SI{343}{\pascal} on inhalation and \SI{245}{\pascal} on exhalation (at flow rates of \SI{85}{\litre\per\minute}) \cite{kim2015,N95std}.
With a fixed limit to $\Delta p$, there are really only two factors that we can vary: the particle collection efficiency of a single fibre, $\lambda$, and the mask geometry through $\alpha$.
In practice, the quality factor $q$ can be optimised by varying the geometric parameters $d_f$ and $\alpha$ (and thus implicitly $\lambda$) by \eg combining layers of different materials.
The resulting efficiency from combining fabric layers has been explored extensively in experiments in \Refscite{konda2020,wangData2020}.

For spatially heterogeneous masks (woven or knitted) \eqref{eq:deltap} no longer applies.
However, from mechanical considerations the pressure drop must be independent of the path through the mask which allows us to treat this more general case.
We will consider the effect this has on the flow through woven materials illustrated in \Figref{fig:schematic}.
Specifically, we consider the inter-yarn pore regions shown in \Figref{fig:schematic}(a-c).
The pores are seen as the light regions under bright-field microscopy in \Figref{fig:schematic}(a), however SEM (\Figref{fig:schematic}(b)) reveals that these pores are \emph{not} empty and so droplet capture can still occur there.
However, these pores contain considerably fewer fibres than inside the yarns so the flow is faster there.

If $U_0$ corresponds to the average flow speed through the entire fabric (constrained by the breathing rate), then we generally expect to find $U_f \ll U_0 \ll U_p$ where $U_f$ and $U_p$ are respectively the average flow speeds through the dense yarn and sparse inter-yarn pore regions.
Typical flow speeds can be estimated by inserting $U_f$ into \eqref{eq:deltap} and equating the pressure drop with that expected through the inter-yarn pores assuming Poiseuille flow.
This yields a relationship between $U_f$ and $U_p$ in terms of the pore area fraction
\begin{equation}\label{eq:pore-area-fraction}
  \kappa = \frac{g_y g_z}{l_y l_z}.
\end{equation}
For typical values of $\kappa$ we find that $\gtrsim \SI{99}{\percent}$ of the flow is expected to go through the pore region, and the \emph{average} flow inside the pore is approximately
\begin{equation}\label{eq:pore-flow}
  U_p \simeq \frac{U_0}{\kappa}.
\end{equation}
This is related to the longstanding `stagnant core problem' of laundry detergency \cite{shin2018}.

\subsection{Extending filtration to woven and knitted materials}
\label{sec:woven}
\subsubsection{Zeroth-order approximation: ignoring pores}

As a zeroth-order approximation to modelling spatially heterogeneous fabrics, we treat them as an effective homogeneous (nonwoven) medium.
We assign each fabric an average quantity $\langle \alpha \rangle$ and $\langle L_x \rangle$, obtained by averaging over the fabric unit cell shown in \Figref{fig:schematic}(a).
\Figref{fig:schematic}(b) shows how yarns elastically deform to have stadium cross-sections where they interlock \cite{kemp1958}, which we approximate as a rectangular cross-section to simplify the averaging procedure.
Thus, the local thickness of the fabric simply equals the sum of diameters of any yarns present while traversing the unit cell in \Figref{fig:schematic}(c); consequently we take the thickness to be zero in the pore region and assign $L_x$ as in \Figref{fig:schematic}(d) where there are yarns:
\begin{itemize}
  \item $L_x = D_y + D_z$ in the four corner regions of the unit cell, occupying a total area $D_y D_z$.
  \item $L_x = D_y$ or $D_z$ in the rectangular regions where there is a single yarn, with areas $g_z D_y$ and $g_y D_z$.
\end{itemize}
$D_y, D_z \gg d_f$ are the thicknesses of the warp and weft yarns (\cf \Figref{fig:schematic}), which we obtained for our sample fabrics in section \ref{sec:experiments} and \citeauthor{zangmeister2020} state these for their fabrics and summarised in table~\ref{tab:zangmeister-cotton}.
This gives the average thickness as
\begin{equation}\label{eq:average-thickness}
  \langle L_x \rangle = \frac{g_y D_z^2 + g_z D_y^2 + (D_y + D_z) D_y D_z}{l_y l_z}. \\
\end{equation}
The average volume fraction $\langle \alpha \rangle$ is then obtained from \eqref{eq:volume-fraction} by combining $\langle L_x \rangle$ with the fabric weight and the bulk density of the material.
By inserting these spatially averaged parameters into \eqref{eq:total-penetration} we can treat a woven fabric as an effectively homogeneous (nonwoven) one.
We thus assume an average flow of $U_0$ through this effective medium in this zeroth-order approximation.

We compare this approximation (dashed line) to literature experimental data for several plain-woven cotton fabrics considered in \citeauthor{zangmeister2020} in \Figref{fig:literature-compare}(b-f).
The agreement with the literature data is poor for small particles, but improves approaching larger particle sizes of $d_p \sim \SI{1}{\um}$.
The smallest particles are unlikely to contain even a single virion, however the poor agreement causes us to overestimate the efficiency in the intermediate size regime so it is worthwhile to improve on this approximation.
We consider the sources of disagreement below and attempt to refine the model.

\begin{table}[t]
  \begin{center}
  \begin{ruledtabular}
  \begin{tabular}{ccccccc}
    fabric
    & $D_x / \si{\milli\metre}$ & $D_y / \si{\milli\metre}$
    & $g_x / \si{\milli\metre}$ & $g_y / \si{\milli\metre}$
    & $\langle L_x \rangle / \si{\milli\metre}$ & $\langle \alpha \rangle$
    \\
    \hline
    1
    & 0.17(1) & 0.15(1)
    & 0.33(6) & 0.33(6)
    & 0.16(2) & 0.48(7)
    \\
    2
    & 0.23(1) & 0.17(1)
    & 0.33(6) & 0.33(6)
    & 0.24(3) & 0.26(4)
    \\
    3
    & 0.25(1) & 0.21(1)
    & 0.33(6) & 0.33(6)
    & 0.32(4) & 0.30(4)
    \\
    4
    & 0.12(1) & 0.13(1)
    & 0.20(2) & 0.25(3)
    & 0.14(2) & 0.65(9)
    \\
    11
    & 0.19(1) & 0.19(1)
    & 0.33(6) & 0.33(6)
    & 0.22(3) & 0.32(4)
  \end{tabular}
  \end{ruledtabular}
  \caption{
    Parameters characterising the plain-woven cotton fabrics considered in \Refcite{zangmeister2020}, with estimated last-digit uncertainties given in parentheses.
    These were estimated from the measurements given in the SM of \Refcite{zangmeister2020}, together with equations \eqref{eq:volume-fraction} and \eqref{eq:average-thickness} for $\langle L_x \rangle$ and $\langle \alpha \rangle$ as described in the text.
  }
  \label{tab:zangmeister-cotton}
  \end{center}
\end{table}

\subsubsection{Correction for pores}

In the section \ref{sec:pressure-drop} we found that most of the flow is expected to go through the inter-yarn pores in textiled materials.
Consequently, compared to flow through a homogeneous material: (i) the effective fibre density will be reduced and (ii) the typical flow speed will be increased.

Effect (i) generically lowers the collection efficiency as there are fewer fibres to collect particles, whereas the effect of (ii) depends on the collection mechanism.
Collection by inertia (impaction) is enhanced by increasing the flow speed, opposing the effect from an effectively reduced fibre density.
After cancellation we thus expect the resulting change in efficiency to be small, and so we do not correct this collection mechanism.
However, the efficiency of collection by diffusion decreases with increasing flow speed, reinforcing effect (i), which is potentially significant.

We attempt to correct the efficiency of filtration by diffusion by replacing $U_0$ with the approximate pore flow speed \eqref{eq:pore-flow} in our calculated P\'eclet number \eqref{eq:peclet}.
We estimate the pore area fraction $\kappa$ using \eqref{eq:pore-area-fraction} with the yarn parameters given in \Refcite{zangmeister2020}.
When we use this flow speed in the expression for diffusion efficiency (SM \ref{sec:diffusion}), we obtain a final filtration efficiency that more closely matches the experimental data of \citeauthor{zangmeister2020} in \Figref{fig:literature-compare}(b-f).
While the precise data of \citeauthor{zangmeister2020} does not extend into the micron regime, the correct position of the minima in \Figref{fig:literature-compare}(b-f) and the trend towards increasing efficiency approaching $\SI{1}{\um}$ (especially in \Figref{fig:literature-compare}(e)) indicates that leaving inertia uncorrected is reasonable.

Considerable variation from fabric to fabric was reported in \Refscite{zangmeister2020,konda2020,wangData2020}, some of which is seen in \Figref{fig:literature-compare}(b-f).
For example, the fabric in \Figref{fig:literature-compare}(b) is roughly equivalent to a surgical mask whereas the fabric in \Figref{fig:literature-compare}(e) considerably outperforms surgical masks.
Conversely, the fabric in \Figref{fig:literature-compare}(f) performs very poorly; \citeauthor{zangmeister2020} writes that this fabric ``had visually open weave structures compared to all other fabrics analyzed'' (\ie $g_x$ and $g_y$ are large) suggesting that the fabric is a poor filter from a combination of having a low thread count and thin yarns.
The biggest difference we can see between the fabrics in panels (e) and (f) is that (e) has a significantly larger fibre density, as measured through $\langle \alpha \rangle$ (\cf table \ref{tab:zangmeister-cotton}).

\begin{figure}
  \centering
  \includegraphics[width=0.9\linewidth]{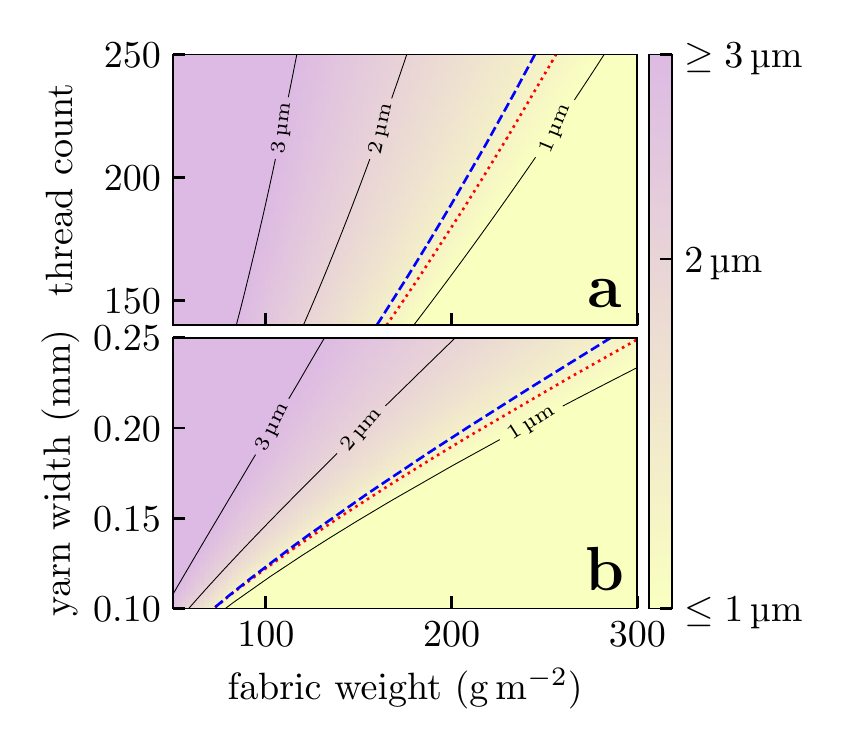}
  \caption{
    (colour online)
    Particle sizes $d_p$ above which woven masks achieve $\ge \SI{95}{\percent}$ filtration.
    Here, we consider 3 layers of identical plain-woven cotton fabrics with (a) fixed yarn widths of \SI{0.2}{\milli\metre} and (b) thread counts of 200.
    For reference we show lines of surgical mask equivalents (blue dashed line) and where the pressure drop across the mask exceeds the \SI{245}{\pascal} threshold set by NIST (red dotted line)\cite{kim2015,N95std}.
    We assume identical warp and weft yarns in these calculations.
    The thread count and fabric weights refer to the properties of individual layers rather than the final multi-layered structure.
  }
  \label{fig:n95-threshold}
\end{figure}

While our model is clearly approximate, it allows us to explore a much wider range of parameters than is typical of experiments to determine the key parameters for effective masks.
In \Figref{fig:n95-threshold} we show how filtration efficiency is expected to be more strongly influenced by the fabric weight than the thread count or yarn sizes in woven fabrics.
The fabric weight is influenced by the thread count, but also by the details of the fabric pattern, the yarn ``crimp'' (\ie how meandering the yarn is in \Figref{fig:schematic}(d)) and the structure of the yarns themselves (\ie how many fibres protrude from the central core).
All else being equal, increasing the fabric weight corresponds to an increased $\langle \alpha \rangle$: this may indicate that the inter-yarn pores are more populated with fibres and gives some crude indication of the fabric's 3d structure.
This is broadly consistent with the explanations proposed by \citeauthor{zangmeister2020} for their best performing fabrics.

\section{Discussion and conclusions}
\label{sec:conclusion}

Masks and face-coverings affect two of the steps in the transmission of a respiratory infection such as COVID-19.
These are exhalation from an infected person, and inhalation by a susceptible person.
Mask effectiveness is not independent of other aspects of transmission, for example, mask efficiency is highest for droplets so large they sediment rapidly.
Sedimentation and aerosol dilution play crucial roles at large physical separations and so mask-wearing is not a substitute for physical distancing.

The basic physics of filtration by fibrous filters, means that filtering out particles of diameter $\gtrsim \SI{3}{\micro\metre}$ is straightforward to achieve in standard fabrics.
Moreover, some fabrics are expect to effectively filter $\ge \SI{95}{\percent}$ particles of diameter $\sim \SI{1}{\micro\metre}$, which is comparable to surgical masks; an example is the first woven cotton fabric studied in \Refcite{zangmeister2020} and shown in \Figref{fig:literature-compare}(b).
Our model makes austere assumptions, so further experiments would be required to refine the parameter range over which these are equivalent.
In particular, the fibre density must be characterised in the inter-yarn pores where most of the air flows through.

For fibres of typical diameters of order $\mathcal{O}(\SI{10}{\micro\metre})$, the Stokes number is of order one or more, and so droplets of this size cannot follow the air streamlines faithfully.
They then deviate from the path of the air flowing through the mask, and so collide with the fibres and are filtered out.
However, filtering out sub-micrometre droplets is much harder as these faithfully track the streamlines of air flowing through the mask.
Without introducing electrostatic interactions, which feature in common fabrics only to a very limited extent it is hard to see how to reliably filter out droplets in this size range.
The sharp cross-over leading to efficient filtration of particles \SIrange{1}{3}{\micro\metre} in diameter emerges from an underlying dynamical transition that was first studied in \Refcite{araujo2006}, and so we expect this to be a robust result.

Even masks made from simple cotton fabrics are predicted to reduce transmission of respiratory viruses, unless transmission is dominated by sub-micrometre droplets.
As masks are cheap, and wearing a mask is a relatively minor inconvenience compared to contracting SARS-CoV-2, recommending mask use is a simple way to reduce transmission.
A simple face covering will never completely eliminate transmission, as some virus-laden droplets will always bypass it.
However, unless transmission is dominated by sub-micrometre droplets, mask use should suppress onwards transmission of the virus.
To the best of our knowledge, sub-micrometre droplets are highly unlikely to carry significant viral loads \cite{freitag2020,robinsonPanic2020}.

Rather than mandating medical-grade PPE, policy makers could pursue a strategy of improving the quality of cloth masks worn in community settings.
Our theoretical model enables the systematic exploration of the mask parameters, which provides a route to optimise mask performance.
We have shown that \emph{under ideal conditions} cloth masks can be optimised to perform as well as surgical masks.
However, the practical performance of any particular mask (cloth or surgical) will crucially depend on the quality of the face seal \cite{freitag2020,duncan2020}.
Practical guidance on reducing leakage would therefore be required to pursue this strategy.
For example, \citeauthor{duncan2020} \cite{duncan2020} found that surgical masks sealed via tie straps offered better face seal than ear loops.

The limited data available on face seal suggests the leakage of a \emph{single} mask is typically around $\sim$\SIrange{25}{50}{\percent} \cite{rengasamy2014,hossain2020}, corresponding to effectively $\sim$\SIrange{5}{25}{\percent} when both inhaler \emph{and} exhaler are masked.
Even with this leakage we would expect a \SIrange{50}{75}{\percent} reduction in exposure to viral particles larger than $\ge \SI{1}{\um}$ under universal masking, or \SIrange{15}{50}{\percent} for sub-micrometre droplets.
Note that a reduction in basic reproduction number $R$ from $R_0 = 4$ by a conservative $\SI{25}{\percent}$ would prevent $\sim \SI{75}{\percent}$ of cases during one month of exponential growth assuming a case doubling time of 3.5 days \cite{bar-on2020}.

Our calculations relied on the standard models of the physics of filtration by fibrous filters.
These capture the essential physics, but rely on simple, two-dimensional, models.
We have generalised these models to incorporate the polydisperse fibre diameter distributions obtained from SEM experiments, as well as to treat the hierarchical (yarn) structure in woven fabrics in an \emph{ad hoc} fashion.
There is scope for future work to look at fully three-dimensional models, models where droplets do not couple to the flow field just at the centre of mass, and models for the fibre/droplet interaction.

By focusing on filtration we have neglected how the mask intervenes with airflow around the mouth and nose, which can significantly change the location and rate of droplet deposition \cite{xi2020,dbouk2020}.
\citeauthor{xi2020}\cite{xi2020} have found that mask wearing strongly perturbs air flow and hence droplet deposition in the respiratory tract, which implies that the reduction in particles deposited in the respiratory tract will be different from the reduction due to filtration.
The authors of \Refscite{xi2020,dbouk2020} did not consider the size-dependence of filtration efficiency, so combining these approaches is a potential avenue for future work.

\begin{acknowledgements}
The authors wish to thank Lewis Husain, for providing the inspiration for this work, Kate Oliver for helpful discussions on textiles which guided our initial investigation, Patrick Warren for guidance on LB simulations, Mahesh Bandi for making us aware of his ingenious use of a candyfloss maker, as well as Mike Allen, Jens Eggers and Flo Gregson for helpful discussions.
We gratefully acknowledge Daniel Bonn, Patrick Charbonneau, Tanniemola Liverpool, John Russo, Hajime Tanaka, and Patrick Warren for providing valuable comments on the manuscript.

JFR, CPR and JPR wish to thank the Bristol Aerosol COVID-19 group for valuable discussions and feedback on this work.
JFR would like to thank Kirsty Wynne for assistance in debugging the code used in the calculations with the Kuwabara flow field.
The authors would like to thank Jean-Charles Eloi and Judith Mantell of the Wolfson Bioimaging Facility and the Chemical Imaging Facility (EPSRC Grant ``Atoms to Applications'', EP/K035746/1), respectively, for the SEM images and assistance in this work.
\end{acknowledgements}

\section*{Data availability statement}

The code used to do the calculations in this work is available at \Refcite{maskflowGitHub}.
Any additional data that support the findings of this study are available from the corresponding author upon reasonable request.

\section*{Supplementary Material}

See supplementary material for the explicit Kuwabara flow field parameters, details of the Lattice Boltzmann simulations and tests validating the filtration theory, scanning electron microscope images and parameters of fabrics obtained from their analysis, the standard model used for treating diffusion collection efficiency, and the electrostatic potential around cylindrical fibres.

\bibliography{everyDayIsLikeSunday-zotero,everyDayIsLikeSunday-manual}

\clearpage{}

\end{document}